\documentclass{jetpl}
\twocolumn

\usepackage{amsmath,amssymb,epsfig,color,pstricks,bm,graphics,alltt}

\usepackage[colorlinks,plainpages=false,linkcolor=black,urlcolor=blue,citecolor=black,pdfpagemode=UseNone,pdfstartview=FitBH]{hyperref}

\definecolor{MyDarkBlue}{rgb}{0,  0.3,  0.9}
\definecolor{MyDarkBlack}{rgb}{0,  0,  0}
\newcommand \modified[1]{\textcolor{black}{#1}}

\begin{document}

\lat

\title{Electronic structure of novel multiple-band superconductor 
SrPt$_2$As$_2$}

\rtitle{Novel multiple-band superconductor SrPt$_2$As$_2$}

\sodtitle{Electronic structure of novel multiple-band superconductor SrPt$_2$As$_2$}

\author{I.\ A.\ Nekrasov\thanks{E-mail: nekrasov@iep.uran.ru}, M.\ V.\ Sadovskii\thanks{E-mail: sadovski@iep.uran.ru}}

\rauthor{I.\ A.\ Nekrasov, M.\ V.\ Sadovskii}

\sodauthor{Nekrasov, Sadovskii }

\sodauthor{Nekrasov, Sadovskii}

\address{Institute for Electrophysics, Russian Academy of Sciences, 
Ural Branch, Amundsen str. 106,  Ekaterinburg, 620016, Russia}

\dates{November 2010}{*}

\abstract{
We present LDA calculated electronic structure of recently discovered
superconductor SrPt$_2$As$_2$ with T$_c$=5.2K. Despite its chemical 
composition and crystal structure are somehow similar to
FeAs-based high-temperature superconductors, the electronic structure of 
SrPt$_2$As$_2$ is very much different.  
Crystal structure is orthorhombic (or tetragonal if idealized) 
and has layered nature with alternating PtAs$_4$ and AsPt$_4$ tetrahedra 
slabs sandwiched with Sr ions.  The Fermi level is crossed by Pt-5d states  
with rather strong admixture of As-4p states. Fermi surface of
SrPt$_2$As$_2$ is essentially three dimensional, with complicated sheets
corresponding to multiple bands. We compare SrPt$_2$As$_2$ with 1111 and 122 
representatives of FeAs-class of superconductors, as well as with isovalent 
(Ba,Sr)Ni$_2$As$_2$ superconductors. Brief discussion of superconductivity
in SrPt$_2$As$_2$ is also presented. 
}


\maketitle

In 2008 new class of FeAs based high-temperature superconductors 
was discovered \cite{kamihara_08}. Reviews of most of pioneering papers were 
presented in Refs.~\cite{UFN_90,Hoso_09}. Lots of experimental and theoretical 
investigations were published since then and the search for new promising
compounds is continuing. 
Recent report on coexistence of superconductivity (T$_c$=5.2K) and CDW 
by Kudo~$et~al.$\cite{Kudo10} in the new arsenide compound SrPt$_2$As$_2$ 
motivated us to compare its electronic structure with that of different 
representatives of FeAs class of superconductors investigated by us earlier 
\cite{Nekr,Nekr2,Nekr3,Nekr4,Kucinskii10}. We shall see that despite
expectations expressed in Ref. \cite{Kudo10} this system cannot be considered
an analogue of Fe pnictides, though by itself presents an interesting case of
a new multiple band superconductor with very complicated electronic structure
close to the Fermi level. 

As reported in Ref.~\cite{Imre07} SrPt$_2$As$_2$ has orthorhombic structure 
with the space group $Pmmn$. To some extent it reminds CaBe$_2$Ge$_2$ type 
tetragonal crystal structure with the space group $P4/nmm$. The later one is
presented in Fig.~1. 
There are two layers of PtAs$_4$ and AsPt$_4$ tetrahedra in the elementary cell 
(shown in Fig.~1). In one layer square lattice of Pt ions is coordinated with 
As tetrahedra and vice versa in the next layer. Corresponding tetrahedra are 
connected with lines in Fig.~1.  The most similar 122 polymorhic form to 
SrPt$_2$As$_2$ is BaFe$_2$As$_2$ pnictide \cite{rott} where two mirrored 
FeAs$_4$ tetrahedra layers are contained in the elementary cell \cite{Nekr2}.  
Other pnictide systems belong mainly to the space group $P4/nmm$ 
\cite{Kucinskii10}.

In present work we neglect for simplicity of comparison to other pnictides
small atomic position modulations in the PtAs$_4$ layer of SrPt$_2$As$_2$ 
leading to the $Pmmn$ structure detected in Ref.~\cite{Imre07}.  
Thus we transform the real $Pmmn$ structure to an idealized $P4/nmm$ as 
described below. To obtain the tetrahedral $a$ lattice parameter we have 
taken the average of $a$=4.482\AA~ and $b$=4.525\AA~ parameters of orthorhombic 
structure, while $c$=9.869\AA~ parameter was taken the same as in the 
experimentally observed structure. Wyckoff Positions of ions were changed 
from $Pmmn$ to $P4/nmm$ like this:  Sr 
2a(1/4,1/4,0.7469)$\to$2c(1/4,1/4,0.7469); for PtAs$_4$ layer Pt1 
4e(1/4,0.8163,0.9989)$\to$2a(1/4,3/4,0) and As2 
4e(1/4,0.294,0.1263)$\to$2c(1/4,1/4,0.1263); for AsPt$_4$ layer Pt2 
2a(1/4,1/4,0.3817)$\to$2c(1/4,1/4,0.3817) and As1 
2b(1/4,3/4,0.4997)$\to$2b(1/4,3/4,1/2). 
Thus we do not account for alternating half filled 4e positions of Pt1 and As2 
and slightly change appropriate coordinates for the atoms.

\begin{figure}[ht]
\includegraphics[clip=true,width=0.45\textwidth]{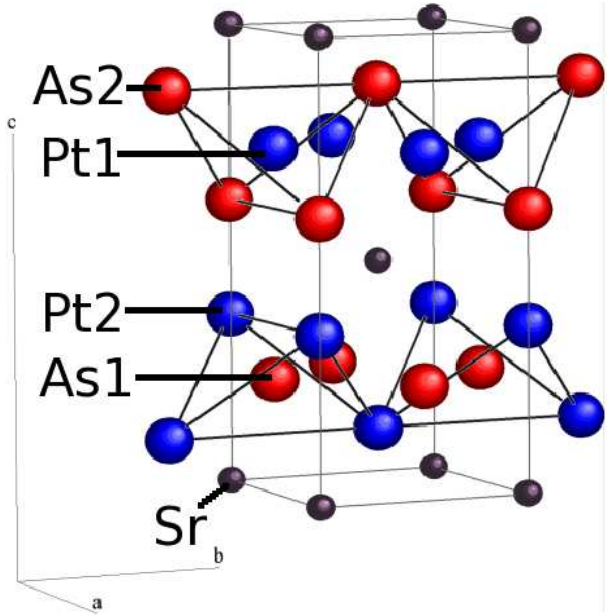}
\label{fig1}
\caption{Fig. 1. Idealized tetragonal crystal structure of SrPt$_2$As$_2$.
} 
\end{figure}

\begin{figure}[hb]
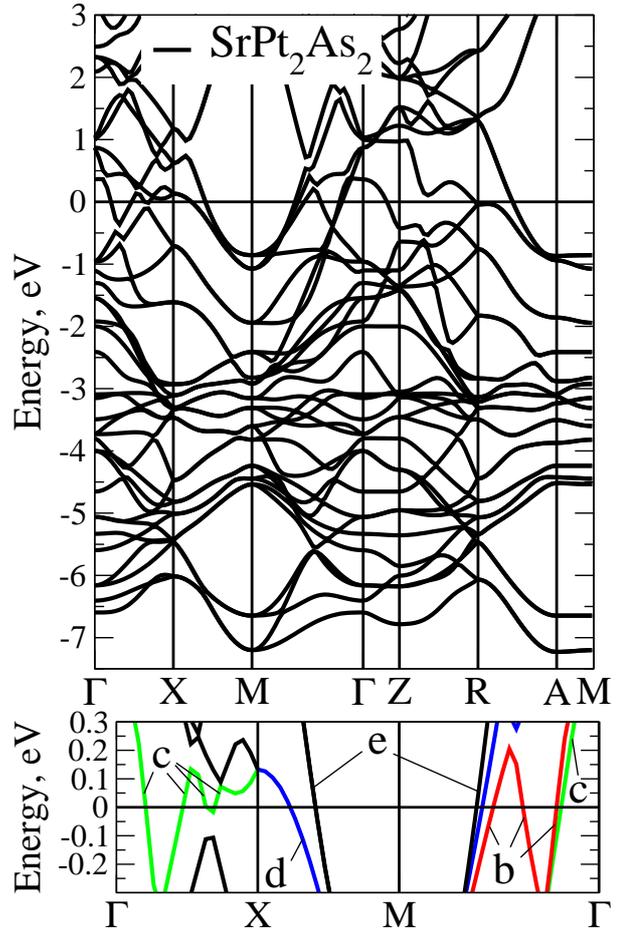

\includegraphics[clip=true,width=0.45\textwidth]{bands_SrPtAs.eps}
\includegraphics[clip=true,width=0.45\textwidth]{bands_ef_SrPtAs.eps}
\caption{Fig. 2. LDA calculated band dispersions of SrPt$_2$As$_2$:
upper panel -- total overview; lower panel -- bands in the $k_z$=0 plane 
in a narrow energy interval around the Fermi level.
\modified{On lower panel letters correspond to Fermi surface sheets shown in Fig. 4.} 
The Fermi level is at zero energy.} 
\end{figure}

Using this idealized SrPt$_2$As$_2$ tetragonal crystal structure we performed 
electronic structure calculations within the linearized muffin-tin orbitals 
method (LMTO)~\cite{LMTO} with default settings.

In Fig.~2 we present band dispersions obtained within LDA calculations 
for SrPt$_2$As$_2$ plotted along high-symmetry Brillouin zone directions. 
Upper panel shows total overview of bands on a large energy scale. 
Lower panel highlights those bands crossing the Fermi level within the $k_z$=0 
plane and in a narrow energy interval relevant to superconductivity. 
\modified{Letters marking} bands on lower panel of Fig. 2 correspond to LDA Fermi 
surface (FS) sheets plotted in Fig.~4 and described below.

One should note that SrPt$_2$As$_2$ bands in general share some 
common features with e.g  1111 FeAs systems \cite{Nekr,Nekr4}, for example 
around M-point. But basically in the vicinity of the Fermi level bands are 
completely different from 1111 and 122 systems. However, the multiple band
nature of electronic spectrum close to the Fermi level is obvious.
There are four band crossings of the Fermi level between
$\Gamma$ and M-points and up to six bands at the Fermi level  
in M-$\Gamma$ direction.

Fig.~3 displays LDA densities of states (DOS) of SrPt$_2$As$_2$. Upper panel 
shows total DOS \modified{(solid line)}. 
Since there are two layers of tetrahedra as discussed above we 
present DOSes for  PtAs$_4$ layer as solid lines \modified{(Pt1 - triangles, As2 -- crosses)}
and for AsPt$_4$ layer as dashed ones
\modified{(Pt2 - squares, As1 -- pluses)}.
Contribution of Pt1-5d states forming square lattice on the Fermi 
level is almost twice larger than of the other states. However, Pt2 and both 
As1,As2 also give considerable contribution to the DOS on $E_F$. 
This distinguishes SrPt$_2$As$_2$ system from Fe pnictides
where As states almost do not appear at the Fermi level \cite{Nekr,Nekr4}.
Most of Pt-5d states are situated much further down from the Fermi level in 
contrast to the pnictides. This can be attributed to the fact that Pt has 
more d-electrons than Fe.  Also Pt-5d states are obviously more extended with 
larger bandwidth than that of Fe-3d.

Lower panel of Fig.~3 shows orbital resolved DOSes for Pt1 and Pt2 5d states. 
Pt1 belongs to  PtAs$_4$ tetrahedron layer and Pt2 to AsPt$_4$ one.
Because of tetragonal symmetry we still can use cubic
notations for the 5d orbitals. For both Pt ions largest contribution
on the Fermi level comes from $x^2-y^2$ orbital (solid line), while other orbitals give
smaller contribution. Situation here differs from that in pnictides 
\cite{Nekr2,Nekr3}, where all of $t_{2g}$ orbitals contribute to DOS at the 
Fermi level. Here as mentioned above most of Pt-5d states lie below the Fermi 
level.

Finally in Fig.~4 we present LDA Fermi surfaces. In Fig.~4a we show all FS 
sheets in the first Brillouin zone. In Figs.~4b--4e we present
different FS sheets separately.
\modified{Letters (b, c, d, e) denoting Fermi sheets correspond to bands marked with the same letters as on lower 
panel of Fig.~2.}
One can see that in total there are six FS sheets. Most of them except Fig.~4e 
are essentially three dimensional. This fact also makes SrPt$_2$As$_2$ 
different from 1111 or 122 pnictides. 
Crossection of the FS at $k_z$=0 is shown in  Fig.~4f with the same \modified{letters} 
coding. Here we also see much more complicated multiple FS sheets picture 
in contrast to Fe pnictides . 

Thus we have in SrPt$_2$As$_2$ the  multiple band structure
with rather complicated multiple sheet FS topology, which is quite different
from that of Fe pnictides. In general, the multiple band system may be also
very complicated from the point of view of possible types of Cooper pairing,
with different energy gaps on different FS sheets, like in FeAs systems
\cite{Gork,KS09}. From the general symmetry analysis \cite{VG,SU} it is known that
in case of tetragonal symmetry and in spin - singlet case we can in principle
observe either the usual isotropic or anisotropic $s$ - wave pairing or 
several types of $d$ - wave pairing. Nothing can be concluded from symmetry
considerations alone on the possibility of $s^{\pm}$ pairing, with e.g.
isotropic gap on different FS sheets changing sign between different sheets,
as it is most probably is in the case of Fe pnictides \cite{UFN_90,KS09}.

As to the value of $T_c$ and gap ratios on different FS sheets in the multiple 
band system these are actually defined by rather complicated interplay of
intraband and interband couplings in Cooper channel, partly determined by
relations between partial DOS'es at these FS sheets \cite{Gork,KS09}. However, we
can make some simple estimates based on elementary BCS approach. Consider
the value of total DOS at the Fermi level N($E_F$) which is 5.6 states/eV/cell
in our calculations. If we calculate corresponding T-linear coefficient of 
the specific heat we obtain $\gamma$=13.1 mJ/mol/K$^2$, which agrees rather 
well with experimental value of 9.7 mJ/mol/K$^2$ \cite{Kudo10}.
Then one can immediately estimate the dimensionless pairing coupling constant 
$\lambda$ value using the BCS expression $T_c=1.14\omega_D e^{-1/\lambda}$,
with Debye frequency $\omega_D$, corresponding to the experiment. 
If we chose $\omega_D$=200K and $T_c$=5.2K in agreement with Ref.~\cite{Kudo10}
for SrPt$_2$As$_2$ we get $\lambda$=0.26. Now we can can estimate $T_c$ values 
for isovalent systems BaNi$_2$As$_2$\cite{Bauer08} and
SrNi$_2$As$_2$\cite{Ronning08} which are also found to be superconductors,
but with much lower experimental superconducting temperatures of
0.7K\cite{Bauer08} and 0.62K\cite{Ronning08}. To do so we performed LDA 
calculations for BaNi and SrNi systems to get appropriate values of N($E_F$)
for these systems. Actually, our results agree well with the previous
LDA calculations of Refs.~\cite{Subedi08,Chen09}. We got N($E_F$)=3.86~states/eV/cell for
BaNi and N($E_F$)=2.81~states/eV/cell for SrNi system. As BCS coupling is always 
proportional to N($E_F$) and assuming the same value of dimensional coupling
in all these systems, we \modified{directly} obtain $\lambda_{BaNi}$=0.18, 
$\lambda_{SrNi}$=0.13 with corresponding $T_c$ values of 0.97K and 0.13K in 
rather good agreement with experiment.  Thus, similar to our work on
pnictides \cite{Kucinskii10} we see that the values of  $T_c$ are well
correlated with the total DOS value at the Fermi level N($E_F$). 

In conclusion, our results show that the novel superconductor SrPt$_2$As$_2$
is essentially a multiple-band system. Despite its crystal structure
similarity to that of 122 pnictides, its electronic structure is quite
different --- more bands cross the Fermi level, Pt-5d states lie at lower 
energies than Fe-3d bands and mostly Pt1-5d $x^2-y^2$  orbital comes to the Fermi level instead of 
all $t_{2g}$ orbitals, contribution of As-4p states at the Fermi level is 
much stronger. Fermi surface has much more complicated multiple sheet 
topology and is definitely three dimensional, which can lead to rather
complicated picture of Cooper pairing. Finally we note that rather non 
monotonous behaviof the DOS near the Fermi level (cf. Fig. 3) suggest the
possibility of significant changes of $T_c$ due to doping.

This work is partly supported by RFBR grant 11-02-00147 and was performed
within the framework of programs of fundamental research of the Russian 
Academy of Sciences (RAS) ``Quantum physics of condensed matter'' 
(09-$\Pi$-2-1009) and of the Physics Division of RAS  ``Strongly correlated 
electrons in solid states'' (09-T-2-1011). IN acknowledges the  grants of the
President of Russia, interdisciplinary UB-SB RAS project.

\onecolumn

\begin{figure}[ht]
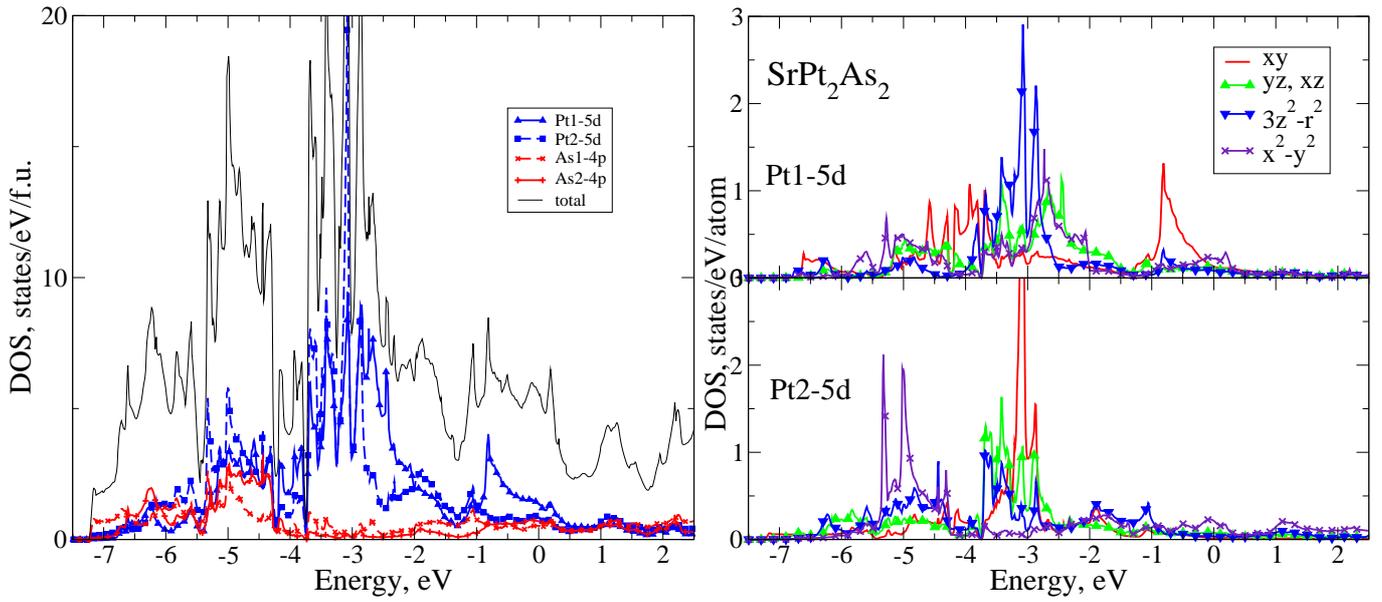

\includegraphics[clip=true,width=0.515\textwidth]{SrPt2As2_dos.eps}
\includegraphics[clip=true,width=0.5\textwidth]{SrPt2As2_dosP.eps}
\caption{Fig. 3. Upper~panel: Densities of states from LDA calculations for 
SrPt$_2$As$_2$.
Solid line -- total DOS;
\modified{solid line with triangles -- Pt1-5d DOS;
dashed line with squares -- Pt2-5d DOS;
solid line with crosses -- As1-4p DOS;
dashed line with pluses As2-4p DOS.}
Lower~panel: DOSes for different orbitals of Pt1 and Pt2 5d shells.
The Fermi level energy is zero.} 
\end{figure}

\begin{figure}[hb]
\includegraphics[clip=true,width=0.3\textwidth]{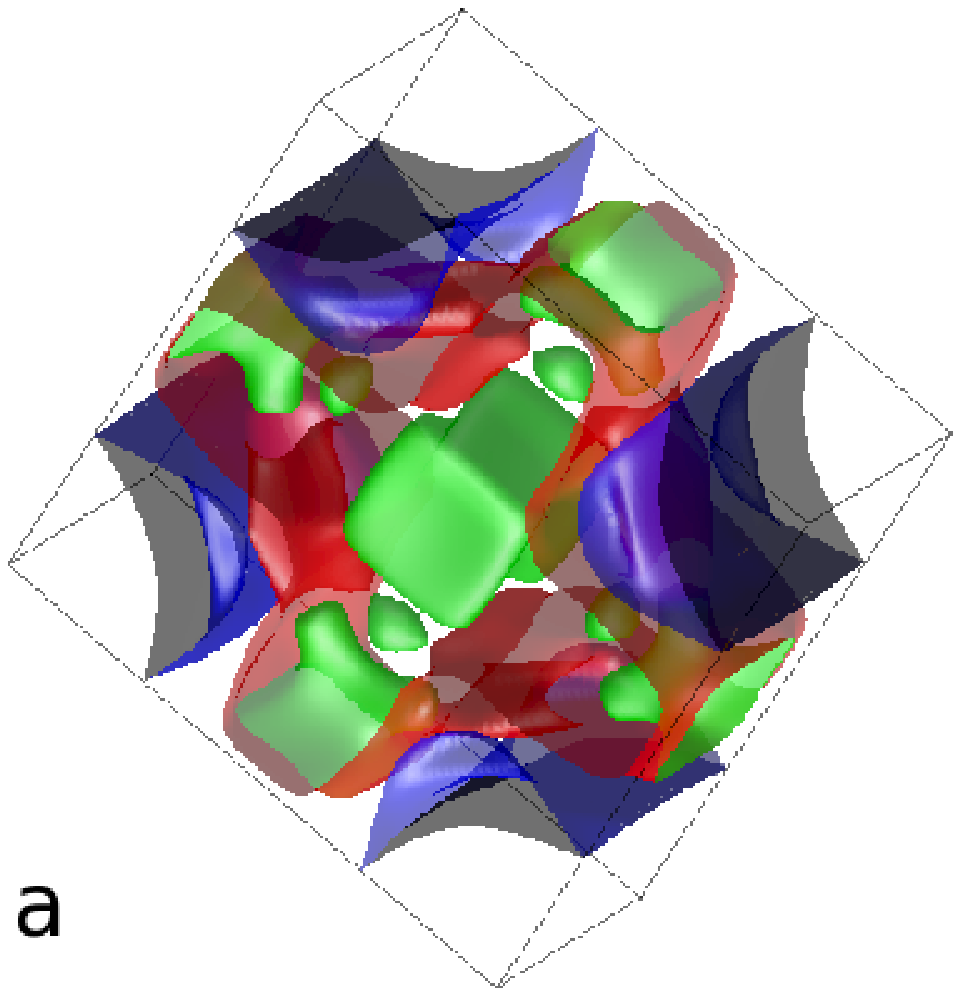}
\includegraphics[clip=true,width=0.3\textwidth]{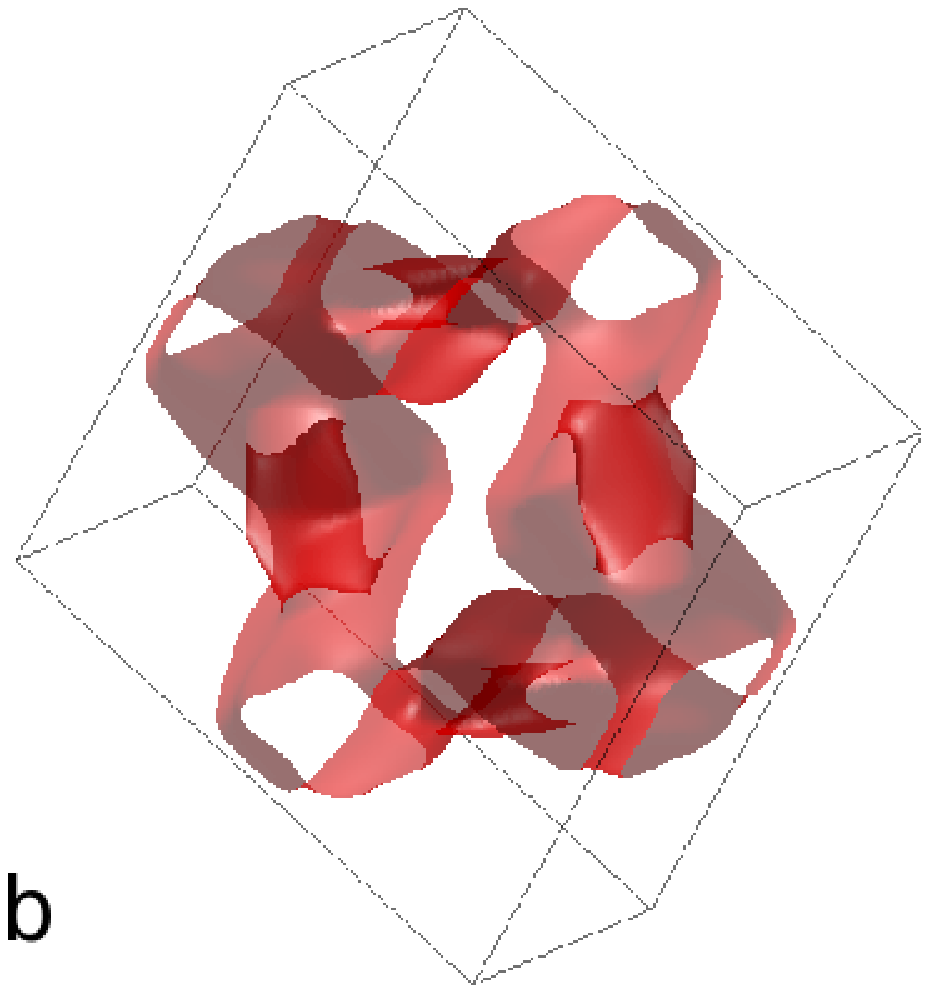}
\includegraphics[clip=true,width=0.3\textwidth]{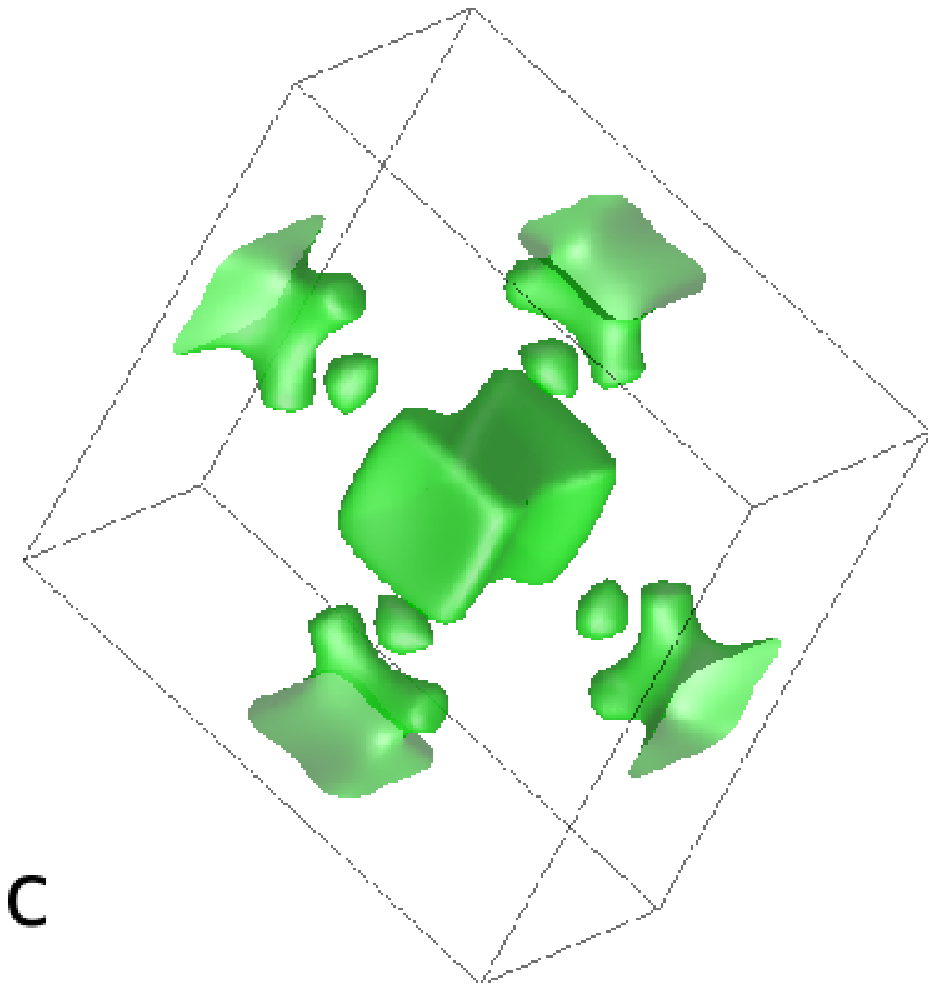}
\includegraphics[clip=true,width=0.3\textwidth]{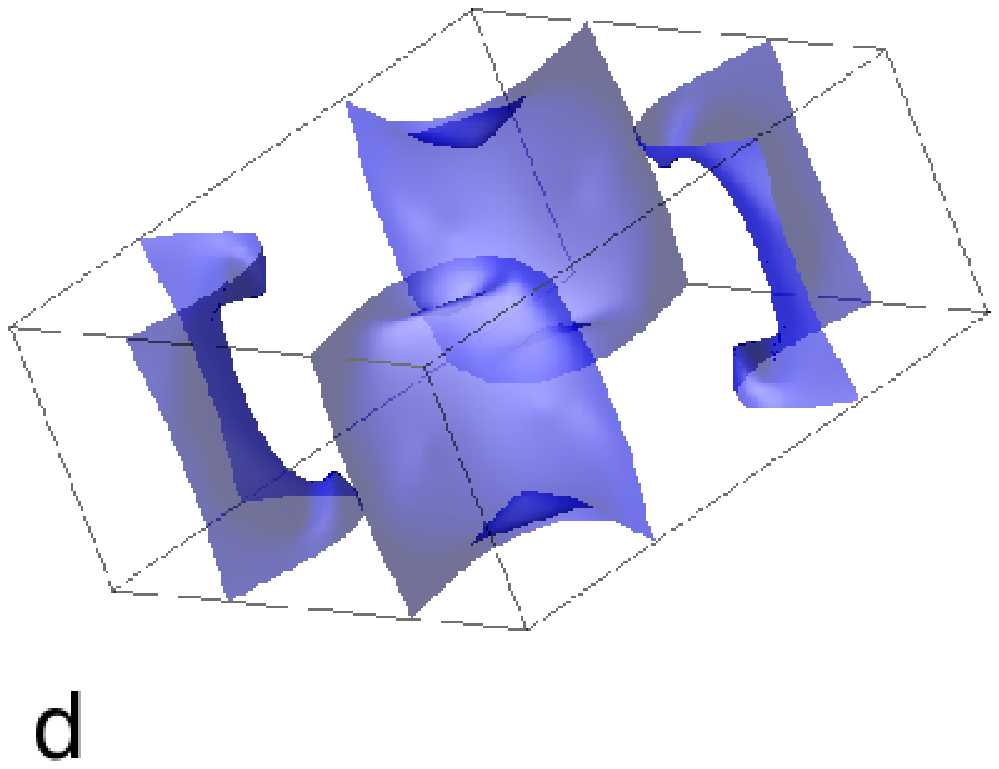}
\hspace{1.cm}\includegraphics[clip=true,width=0.3\textwidth]{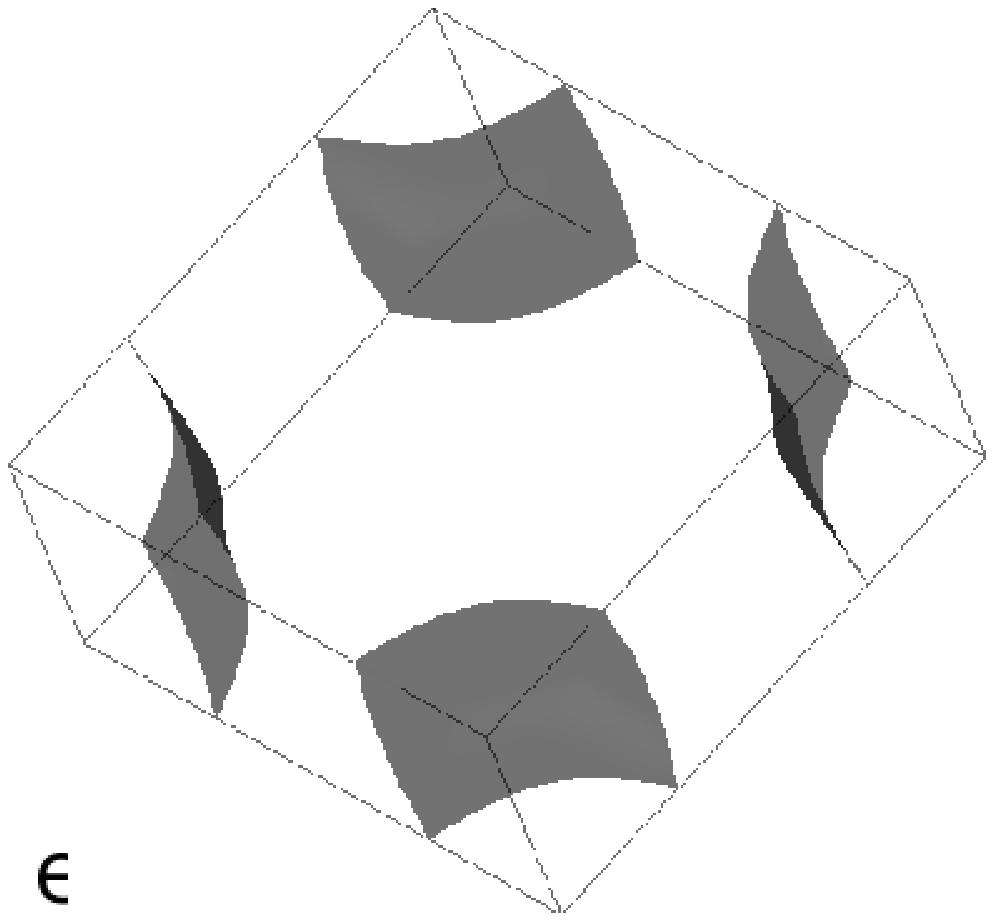}
\hspace{.5cm}\includegraphics[clip=true,width=0.3\textwidth]{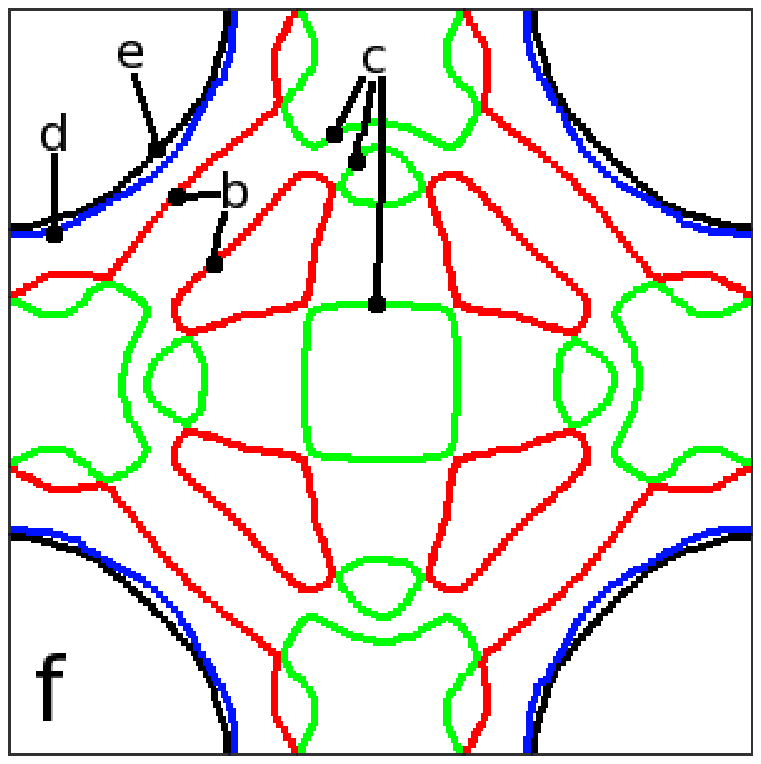}
\caption{Fig. 4. LDA calculated FS for SrPt$_2$As$_2$. 
a -- all FS sheets together;
b,c,d,e -- separate view of each of four FS sheets.  Among these
\modified{(b) is electron-like, (c) is hole-like and  (d) and (e) are electron-like.}
f -- crossection of FS at $k_z$=0  with color coding corresponding to 
Fermi sheets b,c,d and e. } 
\end{figure} 

\twocolumn

\end{document}